\documentstyle[aps]{revtex}
\begin{document}
\draft
\def\lsim{\lower.5ex\hbox{$\; \buildrel < \over \sim \;$}}
\def\gsim{\lower.5ex\hbox{$\; \buildrel > \over \sim \;$}}
%\setcounter{page}
%\begin{center}
%\begin{bf}
\title{ Instantaneous Reflection and Transmission Coefficients and\\
a Special Method to Solve Wave Equation}
%\vskip0.5cm
%\vskip0.35cm
%\end{large}
%\end{bf}
%\end{center}
%\noindent\\
\author  {Banibrata Mukhopadhyay\\}
\address{Theoretical Astrophysics Group\\
S.N. Bose National Centre for Basic Sciences,\\
Block - JD, Sector - III, Salt Lake, Calcutta - 700091, India\\}

%\end{bf}
%\vskip0.5cm
%\noindent\\
%\thanks{Presently in S.N. Bose National Centre for Basic Sciences,
%Block - JD, Sector - III, Salt Lake, Calcutta - 700091, India\\}
%\thanks{Presently in Department of Physics, University of Rochester, Rochester,
%USA\\}
\thanks{e-mail:  bm@boson.bose.res.in  \\}
%\end{center}
%\baselineskip = 12 true pt
\maketitle
\vskip0.3cm
\setcounter{page}{1}
%\noindent{Submitted to appear }
\def\ch{\lower-0.55ex\hbox{--}\kern-0.55em{\lower0.15ex\hbox{$h$}}}
\def\lh{\lower-0.55ex\hbox{--}\kern-0.55em{\lower0.15ex\hbox{$\lambda$}}}	
%\begin{centre}
%{\bf Abstract}
%\end{center}
%\vskip0.3cm
%\noindent
%{\it Received ...., accepted ........}\\
%\vskip0.3cm
%\noindent

\begin{abstract}

{\bf Abstract\hskip0.5cm :} People are familiar with quantum 
mechanical reflection and transmission
coefficient. In all those cases corresponding potentials are usually 
assumed as of constant height and depth. For the cases of varying potential,
corresponding reflection and transmission
coefficients can be found out using WKB approximation method.
But due to change of barrier height,
reflection and transmission coefficients should be changed from point to point.
Here we show the analytical expressions of 
the instantaneous reflection and transmission coefficients.  
Here as if we apply the WKB approximation at each point, so we call it as
Instantaneous WKB method or IWKB method.
Once we know the forward and backward wave amplitudes we can find out
corresponding wave function by calculating {\it Eiconal}. 
For the case of analytically complicated potential 
corresponding differential equation seems to be unsolvable analytically.
If the potentials are well behaved
then obviously these could be replaced by
functions of simple expression of same behaviour
which can be integrated analytically and the equation is
now possible to solve analytically.

\end{abstract}

\vskip2cm
\section*{I. INTRODUCTION}

In the cases of quantum mechanical barrier problem reflection and
transmission coefficients can be calculated. In those cases with the
interaction of the potential field, one fraction of incident
particle (or wave) is reflected back to the infinity (reflection
coefficient) and other fraction transmits into other side or may tunnel
through the barrier (transmission coefficient). If the barrier is
of constant height and depth, the reflection and transmission
coefficient can be easily calculated [1]. In the case of varying potential,
from the far away of potential field, reflection and transmission
coefficient can be calculated using WKB approximation [1,2]. In this
present paper we will show how the reflection and transmission of the
particle changes in the cases of varying potential point to point. Due
to variation of potential, in each point particle should feel different
potential field so the fraction of particle (or wave) which reflects and transmits
will be changing at each point. Here we will calculate the analytical space
dependent expression of reflection and transmission coefficients which
indicates transmission and reflection of the particle at any arbitrary
point with respect to the transmission of the particle at the immediate
previous point. So the local values of incident and reflected wave
amplitude can be found. With this, we shall present the solution of the differential
equation (i.e. Schr\"odinger equation), containing that
space dependent reflected and transmitted amplitude.

Secondly, in our daily problems of physics and mathematics
there are several differential
equations which can not be solved analytically.
If we want to solve those equations
we need numerical methods. If, however, the nature of the coefficients
of derivative terms or inhomogeneous term namely factor, due to
presence of those factors
corresponding differential equation to be unsolvable, are well
behaved then using the methods which will be explained here the
differential equation can be solved analytically using WKB approximation
method. Actual reason of non-analyticity is
that corresponding factors can not be integrated analytically.

We can consider, a particle is moving in a potential field whose
analytical form is complicated. Obviously corresponding
Schr\"odinger equation will be non-analytical. Although the reflection and
transmission coefficients for that particle can be calculated
but the wave function can not be analytically calculated exactly, 
because for this job
we need to integrate analytically the wave vector $k$, which includes
the complicated potential.

In next section we will discuss how the space dependent reflection and 
transmission coefficient can be calculated for the case of varying
potential. In \S 3, we will establish approximate analytical solution of 
a Schr\"odinger equation with the presence of potential with
complicated analytical form. 
In \S 4 we will illustrate an example where our method
can be applied for finding solution and reflection-transmission coefficients.
In \S 5 we will make our conclusions.

\section*{II. Method to find local reflection and transmission coefficients}

We consider an one dimensional Schr\"odinger like differential equation as
$$
\frac{d^2y}{dx^2}+k^2(x)y=0
\eqno{(1)}
$$
where,
$k^2(x)=E-V(x)$, $E=$ total energy, $V(x)=$ potential energy.
\hskip6.5cm(2)  

Here, the independent variable $x$ is varying from $-\infty$ to $\infty$,
i.e., nothing but cartesian co-ordinate. 

Using WKB method the solution of this equation is [1,2]
$$
y(x)=\frac{A}{\sqrt{k}}exp(iu)+\frac{B}{\sqrt{k}}exp(-iu)
\eqno{(3)}
$$
where,
$$
u=\int{k(x) dx},
\eqno{(4)}
$$ 
$A$ and $B$ are constants of integration.

The reflection and transmission coefficients calculated by WKB method means
the coefficients are calculated at a certain point, that means from a certain point
what fraction of particle is transmitted inside and what other fraction
is reflected outside. If we calculate this transmission and reflection
of the particle from different points, results should be different. So
the probability of transmission and reflection of the particle from point
to point will be different for the case of varying potential. So the 
constants $A$ and $B$ should be changed from point to point due to change
of effective potential which is felt by the particle. Here we will
give an analytical space dependent expression for transmission and 
reflection coefficient by which one can find, how the reflection and transmission
probability change from point to point for the case of varying 
potential. 

It is assumed that far away from the potential field (at $x=\infty$), the potential
barrier height is almost constant, so the reflection and transmission
coefficients are almost same. In those regions we safely can choose
$A$ and $B$ as pure constants. Again, all along the sum of transmission
and reflection coefficients should be unity, so far away from the potential
field the relations between $A$ and $B$ are,
$$
A=B+c (c={\rm constant}),
\eqno{(5)}
$$
$$
A^2+B^2=k.
\eqno{(6)}
$$ 
It is very easy to check from incident and reflected current that
transmission and reflection coefficients are $\frac{A^2}{k}$ and
$\frac{B^2}{k}$ respectively.

Solving these two equation we get,
$$
A(x)=\frac{c}{2}+\frac{\sqrt{2k-c^2}}{2}
\eqno{(7a)}
$$
$$
B(x)=-\frac{c}{2}+\frac{\sqrt{2k-c^2}}{2}
\eqno{(7b)}
$$
These $A(x)$ and $B(x)$ are very slowly varying functions 
at large $x$. These are
indicating, at far away from the potential barrier (where potential
is varying in very slow manner), how constants $A$ and $B$ are
slowly changing. So from these slowly varying $A$ and $B$, one can find out 
how transmission and reflection coefficients are varying at far distance.
Now, constant $c$ can be calculated from the boundary
conditions. From pure WKB approximation constant $c$ can be calculated.
From WKB method, far away from the potential field reflection and transmission
coefficients and corresponding reflected and transmitted amplitude can
be calculated. So constant $c$ is immediately determined. 
Obviously, at $x=-\infty$ these expressions of $A$ and $B$ will not be valid 
because of simplified assumptions. At the minimum value of $x$, i.e., 
$x=-\infty$, we have to put another condition.

It is clear that the potential field, which is felt by the particle, 
extended in between $-\infty$ to $\infty$ and assumed that at 
$x=-\infty$ barrier height goes down to zero or constant value i.e., 
the potential is asymptotically flat. So the particle which 
reaches at $x=-\infty$, should feel free. As a result,
there transmission should be $100\%$
and corresponding reflection is zero (hereafter, this will be called as
inner boundary condition). By introducing this inner boundary condition
in eqn. (7) we get,
$$
C(x)=c_1+A(x)= c_1+\frac{c}{2}+\frac{\sqrt{2k-c^2}}{2},
\eqno{(8a)}
$$
$$
D(x)=c_2+B(x)= c_2-\frac{c}{2}+\frac{\sqrt{2k-c^2}}{2}.
\eqno{(8b)}
$$
Here, constants $c_1$ and $c_2$ are introduced to modify the reflection and
transmission coefficients according to the inner boundary condition. 

Here, one necessary condition is the sum of reflection and transmission
coefficients should be one. By modifying the coefficients, it is seen that
$$
C^2(x)+D^2(x)= \left(c_{1} + \frac{c}{2}\right)^2 + \left(c_{2} - \frac{c}{2}\right)^2
+ (c_{1} + c_{2})\sqrt{2k - c^2} + \frac{(2k - c^2)}{2} = h(x) [say] 
=\left[\frac{h(x)}{k(x)}\right]k(x).
\eqno{(9)}
$$
So, it is advisable to choose the modified coefficients of the wave function
as follows:
$$
a(x) = \frac{C(x)}{\sqrt{h/k}}
\eqno{(10a)}$$
$$b(x) = \frac{D(x)}{\sqrt{h/k}}
\eqno{(10b)}
$$
so that
$$
a^2(x) + b^2(x) = k(x)
\eqno{(11)}
$$
and the transmission and reflection coefficients are $a^2(x) \over k(x)$ and
$b^2(x) \over k(x)$ respectively which are explicitly written as:
$$
T(x) = \frac{(c_1 + \frac{c}{2})}{h(x)} \left(c_1 + \frac{c}{2} +
\sqrt{2 k(x) - c^2}\right) + \frac{2 k(x) - c^2}{4 h(x)}
\eqno {(12a)}
$$
$$
R(x) = \frac{(c_2 - \frac{c}{2})}{h(x)} \left(c_2 - \frac{c}{2} +
\sqrt{2 k(x) - c^2}\right) + \frac{2 k(x) - c^2}{4 h(x)}.
\eqno{(12b)}
$$
Now we describe, how to determine the constants $c_1$ and $c_2$.
First of all to satisfy $B(x=-\infty)=0$ (inner boundary condition) $c_2$
is determined and fixed. Then introducing this $c_2$ in $h(x)$ and
equating the expression of modified reflection 
coefficient [$b^2(x)/k(x)$] 
and reflection coefficient valid only at $x=\infty$ [$B^2(x)/k(x)$, 
which is correct at far away
from the black hole]  we get the other constant $c_1$. 
These expressions for transmission and reflection coefficients are valid in
whole region. The space dependent coefficients of incident and reflected
waves, $a(x)$ and $b(x)$ are valid for arbitrary $x$.

So the final form of the solution using WKB method is 
$$
y(x)=\frac{a(x)}{\sqrt{k}}exp(iu)+\frac{b(x)}{\sqrt{k}}exp(-iu).
\eqno{(13)}
$$
Here one important thing is to be noted that, WKB approximation
is valid only if $\frac{1}{k}\frac{dk}{dx} < < k$. So one can
calculate the space dependent reflection and transmission coefficient
only if potential vary in such a manner that $k$ satisfies the above
condition. Otherwise our method to find out space dependent amplitudes
and coefficients is not valid because our method is based on WKB
approximation. Equation (13) is valid at any point. It is indicating
the solution and corresponding reflection and transmission coefficients
for an arbitrary $x$ as if at each point WKB approximation method is
applied instantaneously. So our modified WKB method can be called
as {\it Instantaneous WKB Method} or IWKB method [3]. For the cases, where
$\frac{1}{k}\frac{dk}{dx} < < k$ is not satisfied, the potential can
be replaced by a large number of square steps 
and space dependent reflection and transmission coefficients 
can be found [3]. Here, corresponding problem
is reduced to as simple quantum mechanical barrier problem with multiple
steps where at each step junction the wave functions and its 
derivatives of two separate regions should be continuous. Detail
discussion is in Mukhopadhyay \& Chakrabarti [3].

\section*{III. Solution of the equation}
In the previous section, although we have given the form of the
solution with space dependent incident and reflected coefficients
but the {\it Eiconal} $u$ still yet to be determined. If the analytical
form of the potential and corresponding wave vector $k$ is well
integrable analytically then $u$ can be calculated immediately but
sometimes the form of $k$ may be complicated such that it can not be
integrated analytically then $u$ can not be evaluated unless very 
special cases are chosen.
In this section we want to give an approximate solution of eqn. (1)
using WKB (actually IWKB, as mentioned in last section) 
approximation even if the analytical form of $k$ is not well integrable. 
As we mentioned in INTRODUCTION that if $k^2(x)$ is well behaved then
we can give an analytical form of the eqn. (1) using the following trick.
If the nature of the coefficients of derivative and zeroth order derivative
terms are well behaved but analytically complicated then those analytical
form of the coefficients can be replaced by piecewise continuous
analytical function with simple form and of same behaviour as exact 
complicated functions. After this replacement $k$ will be of simple
form and we can integrate it to find out $u$. Once we know the analytical
form of $u$, we can present the analytical solution 
of the equation. 

In different ways we can replace the complicated functions. 
Any function can be expanded in terms of polynomial. It is seen that
if the corresponding polynomial is of order two then the corresponding
$k$ can be integrated easily. But using one polynomial of order 
two, in the total range, the replacement of the complicated function
can not be done. So in different regions, different coefficients of
the polynomial are chosen i.e., different second order polynomials
are chosen such that at the boundaries (where the polynomials change),
value of the two different polynomials and its derivatives are same.
In this way, by using many number of piecewise polynomials the analytically
complicated well behaved functions can be replaced. 

If in the differential eqn. (1), the analytical form of $V$ and
corresponding $k$ is of non-analytical form (complicated analytical
expression) then potential $V$ can be replaced by piecewise 
continuous polynomial as
$$
V_l(x)=a_l+b_lx+c_lx^2
\eqno{(14)}
$$
and the corresponding $k(x)$ is defined as
$$
k^2_l(x)=(E-a_l)-b_lx-c_lx^2.
\eqno{(15)}
$$
Integrating this we can find out analytical form of $u$ as
$$
u_l(x)=-2c_lk_l-c_l\sqrt{(E-a_l)}log\left|{{k_l-\sqrt{(E-a_l)}}
\over {k_l+\sqrt{(E-a_l)}}}\right| +{\rm constant}
\eqno{(16)}
$$
Where index $l$ in the coefficients of the polynomial is indicating
that in different ranges of $x$ different values of $a$, $b$, $c$ may
be chosen.
This type of replacement also can be done by some other well integrable
functions according to the nature of the complicated functions. In
the next section we will give an example, where, to solve the 
differential equation with the potential of complicated analytical
form, we replace the complicated 
function (potential) by the function of simple analytical form otherthan 
second order polynomials.

In this way we can give the analytical form of the solution of
Schr\"odinger equation like equation (3) even if the analytical
form of the potential
is complicated. Since we have used the WKB approximation method (more
clearly IWKB method),
our solution only will be valid for $\frac{1}{k}\frac{dk}{dx} < < k$.
If the potential vary in such a way that above condition is not
being satisfied then the solution of the Schr\"odinger or Schr\"odinger
like equation is not possible by this method. For those cases we have
to employ some other methods, one of them is indicated in last section 
(by reducing the problem to simple quantum mechanical barrier problem). 
Sometimes for those cases, the Schr\"odinger
equation can be reduced into Bessel equation and solution is possible
by expressing in terms of Airy functions [1].

\section*{IV. illustrative example}

Dirac equation in Kerr geometry can be separated into radial and
angular parts [4-5]. Then corresponding decoupled radial equation on
some transformation of independent variable can be reduced into
Schr\"odinger like equation [5] as 
$$
\frac{d^2Z}{dx^2}+\left(\sigma^2-V(x)\right)=0.
\eqno{(17)}
$$
In comparison with Schr\"odinger equation with $\ch=c=G=1$, we can say
$V(x)$ is nothing but potential and $\sigma^2$ is proportional to
energy of the particle. Here the analytical form of the
potential is complicated as
%\newpage
$$
V={{\Delta^{1 \over 2}(\lh^{2} + m_p^{2} r^{2})^{3/2}} \over {[ \omega^{2}(\lh^{2} + m_p^{2}
r^{2}) + \lh m_p \Delta/2 \sigma]^{2}}}[\Delta^{1 \over 2}(\lh^{2} + m_p^{2} r^{2})^{3/2} +
 ((r-M)(\lh^{2} + m_p^{2} r^{2}) + 3m_p^{2} r \Delta)]
$$

$$
- {{\Delta^{3 \over 2}(\lh^{2} + m_p^{2} r^{2})^{5/2}} \over {[ \omega^{2}(\lh^{2} + m_p^{2}
r^{2}) + \lh m_p \Delta/2 \sigma]^{3}}}[2r(\lh^{2} + m_p^{2} r^{2}) + 2 m_p^{2} \omega^{2} r +
 \lh m_p (r-M)/\sigma] ,
\eqno{(18)}
$$ 
where $a=$ Kerr parameter, $m_p=$ mass of the incident particle, $\sigma=$ frequency of the
incident particle, $M=$ mass of the black hole, $\lh=$ separation constant, $r=$ radial 
co-ordinate, $\Delta=r^2-2Mr+a^2$, $\omega=r^2+a^2+am/\sigma$, $x=f(r)$. 
For details see [5]. 

Now, although the analytical form of the potential is complicated but for particular set
of physical parameter (such as, $a, \sigma, m_p$ etc.) the nature of $V$ is 
well behaved. So we can replace it by suitable peace-wise continuous analytical
function of simple form. We can choose it as second order polynomial as 
explained in previous section. But for convenience we choose here in different
form.

To show the nature of the potential, we choose as an example one set of
physical parameter, such as:

$a= 0.5$; $M=1$; $m_p=0.8$; orbital quantum number, $l=\frac{1}{2}$; 
azimuthal quantum number, $m=-\frac{1}{2}$; $\sigma=0.8$; $\lh = 0.92$ 
from [6].

The simple analytical form which is chosen to replace the complicated form of
the potential for this set of physical parameter is given as
$$
V_l(x)=a_l+b_lexp\left(-\frac{x}{c_l}\right)
\eqno{(19)}
$$
and corresponding $k$ and eiconal $u$ are given as
$$
k_l(x)=\sqrt{(\sigma^2-a_l)-b_lexp(-\frac{x}{c_l})},
\eqno{(20)}
$$
$$
u(x)=-2c_lk_l(x)+c_l\sqrt{(\sigma^2-a_l)}log \left|{{\sqrt{(\sigma^2-a_l)}
+k_l(x)} \over {\sqrt{(\sigma^2-a_l)}-k_l(x)}}\right|+{\rm constant}.
\eqno{(21)}
$$
The behaviour of the functions of eqn. (18) and (19) are same.
In case of (19) in different ranges of $x$ different values of
$a, b, c$ are chosen. 

For these physical parameters, coefficients $a, b, c$ of mapping function in
different ranges of $x$ are given as:

$a = 0$, $b = -0.187354$, $c = -3.75$ for $x$ $-\infty$ to $0$,

$a = 0.603$, $b = 0.415646$, $c = 8.79$ for $x$ $0$ to $30$,

$a = 0.629$, $b = 0.12690038$, $c = 26.3$ for $x$ $30$ to $109$,

$a = 0.63543098$, $b = 0.037193439$, $c = 73.5$ for $x$ $109$ to $208$,

$a = 0.63543098$, $b = 0.2228925$, $c = 45$ for $x$ $208$ to $310$,

etc. 

So finally we can say the analytical form of the solution of the 
Schr\"odinger equation is possible for this set of physical parameter.
Similarly for other sets of physical parameter one can find out 
solution following same method (by mapping the complicated analytical 
form of the potential by the simple form which may be polynomial or
analytical function like eqn. (19) and which are analytically integrable).

The space dependent transmission and reflection coefficients can be 
calculated following the method explained earlier. Now we will calculate
the constants $c, c_1, c_2$ by imposing boundary conditions for the
potential of given set of physical parameter. Using general WKB
approximation method, from the far away of the black hole (say at $x=310$) 
the reflection and transmission coefficients of the particle 
for the potential shown in Fig. 1 can be calculated as $T=0.299$ and $R=0.701$. 
Using these values of the coefficient constant $c$ can be calculated as
$$
c=\sqrt{Tk}-\sqrt{Rk}=-0.0913
\eqno{(22)}
$$
where $k$ is the value of wave number at $x=310$.

%\begin{figure}
%\vbox{
%\vskip -0.5cm
%\hskip 0.0cm
%\centerline{
%\psfig{figure=fig1.ps,height=13truecm,width=13truecm,angle=0}}}
%\vspace{-0.0cm}
%\noindent {\small {\bf Fig. 1} : Variation of potential $V$ (solid curve) 
%with respect to $x$ and total energy $E$ (dashed curve) of particle.}
%\end{figure}
Using this $c$ when we calculate the transmitted and reflected amplitude
from (7a-b) (which are valid at large distance from the black hole) 
obviously it will not satisfy the inner boundary condition. To satisfy 
the inner boundary condition first $c_2$ is introduced to reduce $B(x)$
to zero at $x=-\infty$ as $c_2=-0.6765$. Then as explained in section II, by
matching the reflection amplitudes of eqn. (7b) and (10b) at infinity i.e.,
$$
b(x)=B(x),
\eqno{(23)}
$$
$c_1$ can be evaluated as $c_1=0.1639$.
 
%\begin{figure}
%\vbox{
%\vskip -0.5cm
%\hskip 0.0cm
%\centerline{
%\psfig{figure=fig1.ps,height=13truecm,width=13truecm,angle=0}}}
%\vspace{-0.0cm}
%\noindent {\small {\bf Fig. 2} : Variation of transmission (solid curve) and
%reflection coefficients (dashed curve) with respect to $x$.}
%\end{figure}
Now all the constants are known. In Fig. 2 variation of instantaneous 
reflection and transmission coefficients are shown. It is seen that
as the potential barrier lower down the transmission probability
increases and reflection probability decreases. For the cases of other
set of physical parameters where the potential barrier may be of different
type, using the same method constants can be calculated.
In the final solution with the insertion of the value of constants it can be
checked that WKB approximation is still valid except in the regions
where $\sigma^2$ {\rm{is close to}} $V$ and this 
modified WKB (IWKB) solution 
satisfies the original differential equation if $\frac{1}{k}\frac{dk}{dx} <<k$. 

\section*{V. conclusions}

Here, we show how reflection and transmission coefficients vary for
the particle moving in spatially varying potential field. Actually
we solve the Quantum Mechanical barrier problem, 
where the barrier height changes
point to point. For solution, we introduce modified 
WKB approximation method namely 
instantaneous WKB method i.e., IWKB method where at each point,
instantaneously, the WKB approximation method can be applied and
corresponding reflection and transmission coefficients can be
calculated. Since the potential changes point to point
corresponding reflection and transmission coefficients change.
We also indicate how Schr\"odinger equation is solved analytically
where coefficients of derivative terms are of complicated analytical
form. It is shown that if coefficients are well behaved then the
the analytical solution can be set up by suitably mapping the 
coefficient-functions to functions of simple form. In this
way, any such kind of differential equation can have analytical
form of the solution which could be useful for further study.   
   
\section*{VI. acknowledgment}

It is a great pleasure to the author to thank Prof. Sandip K. Chakrabarti
for many serious discussion regarding this work and support to write this
paper.

\section*{Figure Caption}

\noindent {\small {\bf Fig. 1} : Variation of potential $V$ (solid curve) 
with respect to radial distance $x$ and total energy $E$ 
(dashed curve) of particle.}\\
\noindent {\small {\bf Fig. 2} : Variation of transmission (solid curve) and
reflection coefficients (dashed curve) with respect to radial distance $x$.}

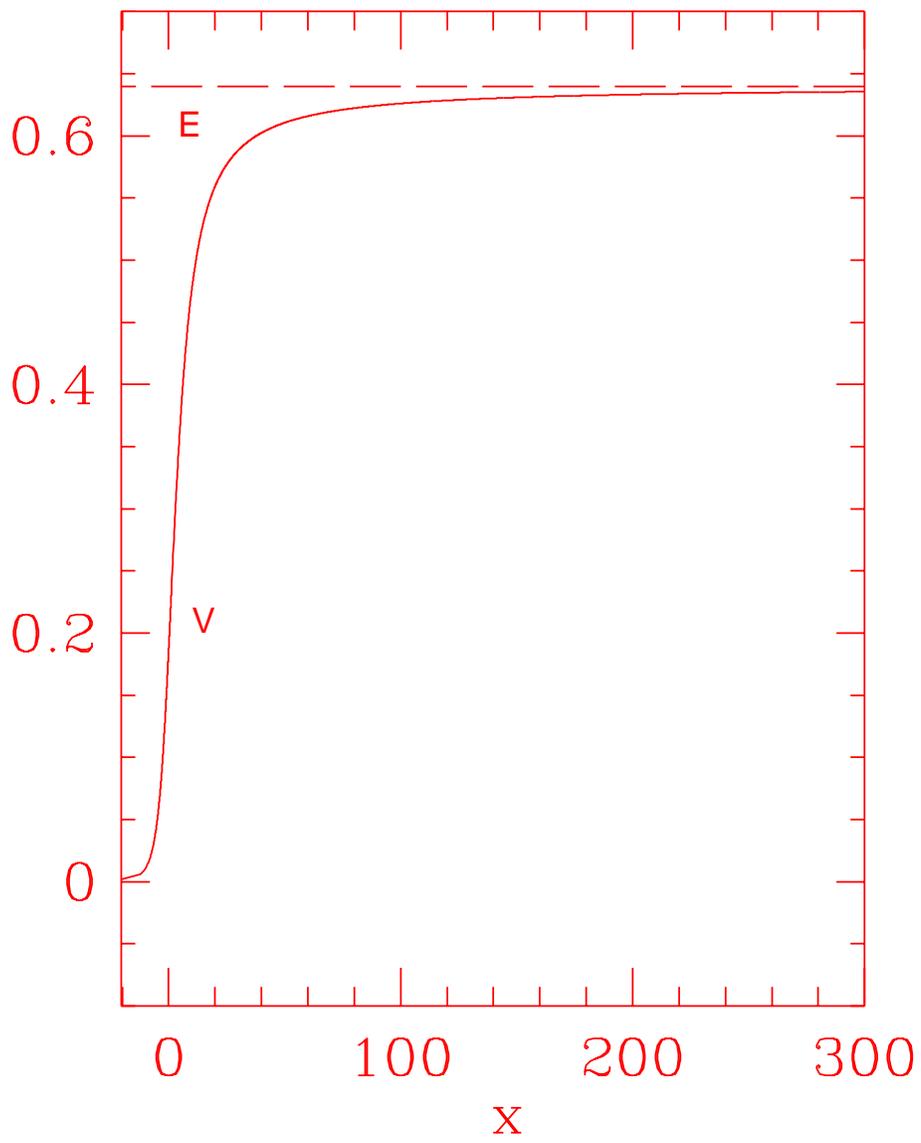
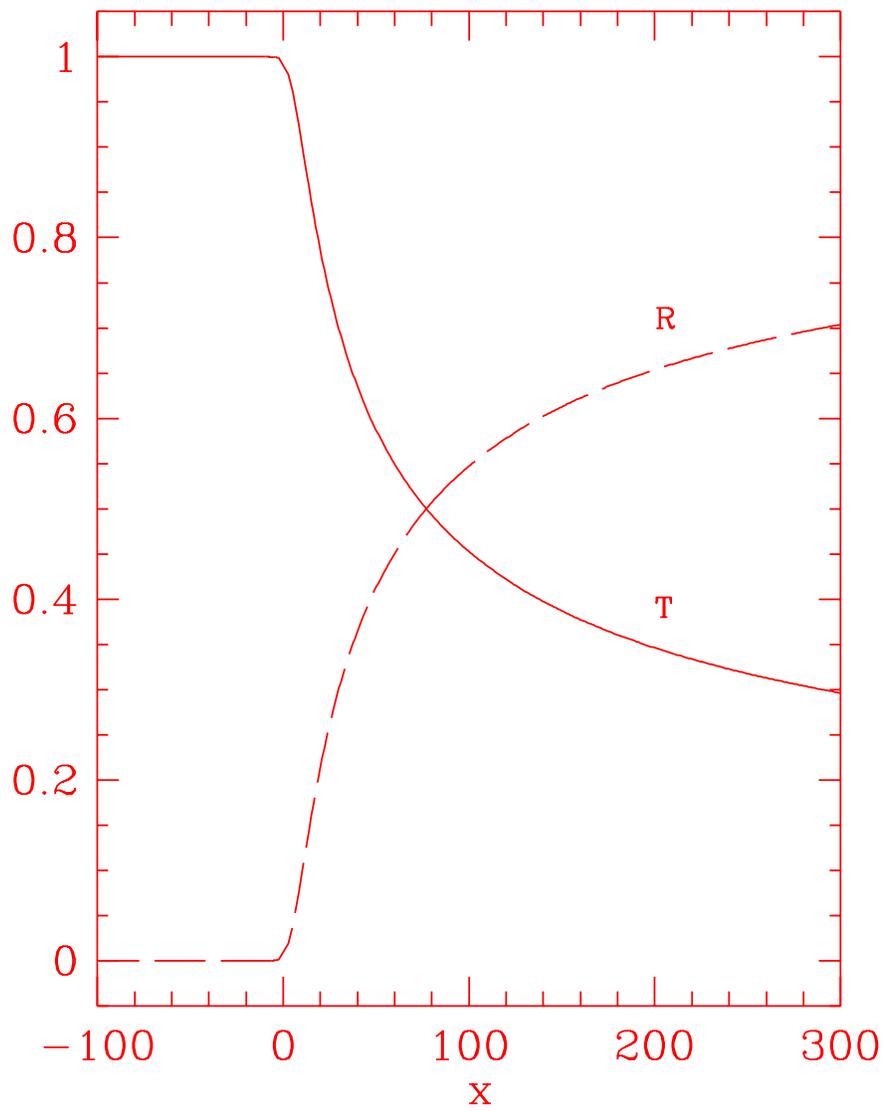
\end{document}